\documentclass[12pt]{article}
\usepackage{amsmath}
\usepackage{amssymb}

\begin{document}

\begin{center}
\textbf{HIGHER DIMENSIONAL GRAVITY AND FARKAS}

\bigskip

\textbf{\ PROPERTY IN ORIENTED MATROID THEORY}

\bigskip

\bigskip

\smallskip\ 

J. A. Nieto$^{\star \ast }$ \footnote{%
nieto@uas.uasnet.mx} and E. A. Le\'{o}n$^{\ast }$ \footnote{%
ealeon@posgrado.cifus.uson.mx}

\smallskip\ 

$^{\star }$\textit{Facultad de Ciencias F\'{\i}sico-Matem\'{a}ticas de la
Universidad Aut\'{o}noma} \textit{de Sinaloa, 80010, Culiac\'{a}n Sinaloa, M%
\'{e}xico.}

$^{\ast }$\textit{Departameto de Investigaci\'{o}n en F\'{\i}sica de la
Universidad de Sonora, Hermosillo Sonora , M\'{e}xico}

\bigskip\ 

\bigskip\ 

\textbf{Abstract}
\end{center}

\noindent We assume gravity in a $d$-dimensional manifold $M$ and consider a
splitting of the form $M=M_{p}\times M_{q}$, with $d=p+q$. The most general
two-block metric associated with $M_{p}$ and $M_{q}$ is used to derive the
corresponding Einstein-Hilbert action $\mathcal{S}$. We focus on the special
case of two distinct conformal factors $\psi $ and $\varphi $ ($\psi $ for
the metric in $M_{p}$ and $\varphi $ for the metric in $M_{q}$), and we
write the action $\mathcal{S}$ in the form $\mathcal{S=S}_{p}\mathcal{+S}%
_{q} $, where $\mathcal{S}_{p}$ and $\mathcal{S}_{q}$ are actions associated
with $M_{p}$ and $M_{q}$, respectively. We show that a simplified action is
obtained precisely when $\psi =\varphi ^{-1}$. In this case, we find that
under the duality transformation $\varphi \leftrightarrow \varphi ^{-1}$,
the action $\mathcal{S}_{p}$ for the $M_{p}$-space or the action $\mathcal{S}%
_{q}$ for the $M_{q}$-space remain invariant, but not both. This result
establishes an analogy between Farkas property in oriented matroid theory
and duality in general relativity. Furthermore, we argue that our approach
can be used in several physical scenarios such as 2t physics and cosmology.

\ \ 

\bigskip\ \bigskip\ 

\bigskip\ 

\bigskip\ 


Pacs numbers: 04.60.-m, 04.65.+e, 11.15.-q, 11.30.Ly

December, 2009\newpage \noindent Duality has been an important concept in
several physical scenarios, including string theory [1], $M$-theory [2],
Matroid theory [3], MacDowell-Mansouri gravity (see Refs. [4]-[5] and
references therein) and cosmological models [6][7]. Here, we explore the
possible relevance of duality in higher dimensional gravity. We first
consider general relativity in a $d-$dimensional manifold $M$ with arbitrary
space-time signature. We then proceed to split such manifold in the form $%
M=M_{p}\times M_{q}$, with $p+q=d$, assuming a two-block diagonal metric.
Considering such splitting, we obtain the Einstein-Hilbert action $\mathcal{S%
}$. For the special case in which the two metrics associated with $M_{p}$
and $M_{q}$ are expressed in terms of the conformal factors $\psi $ and $%
\varphi $, respectively, we show that the action $\mathcal{S}$ can be
written as $\mathcal{S=S}_{p}\mathcal{+S}_{q}$, where $\mathcal{S}_{p}$ and $%
\mathcal{S}_{q}$ are actions associated with $M_{p}$ and $M_{q}$,
respectively. Taking $\psi =\varphi ^{-1}$, the reduced action is analysed.
The key idea is to consider the invariance of such reduced action under the
duality transformation $\varphi \leftrightarrow \varphi ^{-1}$. In this
context, we find that the invariance of $\mathcal{S}$ does not follow, but
rather we discover that, under this duality transformation, the action $%
\mathcal{S}_{p}$ is invariant, provided $p\rightleftarrows q+2$, while $%
\mathcal{S}_{q}$ is not. And conversely, the action $\mathcal{S}_{q}$ is
invariant, provided $p\rightleftarrows q-2$, while $\mathcal{S}_{p}$ is not.
We argue that this result establishes an analogy between Farkas property in
oriented matroid theory [8] and duality in higher dimensional gravity. It is
worth mentioning that the Farkas property has been used as an alternative to
define the concept of oriented matroid [9]. In turn, oriented matroid theory
has been proposed [10]-[11] as the appropriate mathematical framework for
considering duality in several physical scenarios, including p-branes and
M-theory. Therefore, our work may be useful in the analysis of some aspects
of duality in p-branes and M-theory. Furthermore, we argue that our approach
can also be relevant in the context of 2t physics and cosmology.\bigskip

We start the analysis with a metric $\gamma _{AB}$ in a $d-$dimensional
manifold $M$, which will be splitted into a two-blocks metric corresponding
to $M=M_{p}\times M_{q}$. The first block metric, of dimension $p$, will be
denoted by $g_{\mu \nu }$, and the second of dimension $q=d-p$, will be
denoted by $g_{ij}$, where Greek indices ($\alpha $, $\beta $, ...) run from 
$1$ to $p$, lowercase Latin indices ($i$, $j$, ...) from $p+1$ to $d$, and
capital Latin indices ($A$, $B$, ...) from $1$ to $d$. With this
prescription, we have [12][13]:

\begin{equation}
\gamma _{AB}=\left( 
\begin{array}{cc}
g_{\mu \nu }(x,y) & 0 \\ 
0 & g_{ij}(x,y)%
\end{array}%
\right) ,  \label{1}
\end{equation}%
where for consistency, the upper zero $0$ corresponds to a $p\times q$%
-matrix and the lower zero $0$ corresponds to a $q\times p$-matrix. Here, $x$
refers to coordinates in $M_{p}$, while $y$ refers to coordinates in $M_{q}$%
.\bigskip

\noindent As usual, the Riemann tensor is defined in terms of the
Christoffel symbols as%
\begin{equation}
\mathcal{R}_{\ BCD}^{A}=\partial _{C}\Gamma _{DB}^{A}-\partial _{D}\Gamma
_{CB}^{A}+\Gamma _{CF}^{A}\Gamma _{DB}^{F}-\Gamma _{DF}^{A}\Gamma _{CB}^{F},
\label{2}
\end{equation}%
where in turn, the Christoffel symbols are%
\begin{equation}
\Gamma _{BC}^{A}=\frac{1}{2}g^{AF}(g_{BF,C}+g_{CF,B}-g_{BC,F}).  \label{3}
\end{equation}%
From (3), using the distinction of the indices $\mu ,\nu ,...etc.$ and $%
i,j,...etc.$, we find that the Christoffel symbols can be splitted in six
classes:%
\begin{equation}
\begin{array}{ccc}
\Gamma _{\alpha \beta }^{\mu }=\left\{ _{\alpha \beta }^{\mu }\right\} , & 
\Gamma _{\alpha i}^{\mu }=\frac{1}{2}g^{\mu \lambda }g_{\alpha \lambda ,i}\ ,
& \Gamma _{ij}^{\mu }=-\frac{1}{2}g^{\mu \lambda }g_{ij,\lambda }, \\ 
&  &  \\ 
\Gamma _{jk}^{i}=\left\{ _{jk}^{i}\right\} & \Gamma _{j\alpha }^{i}=\frac{1}{%
2}g^{il}g_{jl,\alpha }\ , & \Gamma _{\alpha \beta }^{i}=-\frac{1}{2}%
g^{il}g_{\alpha \beta ,l}.%
\end{array}
\label{4}
\end{equation}%
Here, $\left\{ _{\alpha \beta }^{\mu }\right\} $ and $\left\{
_{jk}^{i}\right\} $ refer to Christoffel symbols in terms of the metric $%
g_{\alpha \beta }$ and $g_{ij}$, respectively.\bigskip

Now, the components of the Riemann tensor can also be splitted in similar
classes. For example:%
\begin{equation}
\mathcal{R}_{\ \nu \alpha \beta }^{\mu }=R_{\ \nu \alpha \beta }^{\mu
}+\Gamma _{\alpha k}^{\mu }\Gamma _{\beta \nu }^{k}-\Gamma _{\beta k}^{\mu
}\Gamma _{\alpha \nu }^{k},  \label{5}
\end{equation}%
where

\begin{equation}
R_{\ \nu \alpha \beta }^{\mu }=\partial _{\alpha }\Gamma _{\beta \nu }^{\mu
}-\partial _{\beta }\Gamma _{\alpha \nu }^{\mu }+\Gamma _{\alpha \lambda
}^{\mu }\Gamma _{\beta \nu }^{\lambda }-\Gamma _{\beta \lambda }^{\mu
}\Gamma _{\alpha \nu }^{\lambda }.  \label{6}
\end{equation}%
Using (4), the relation (5) becomes%
\begin{equation}
\mathcal{R}_{\ \nu \alpha \beta }^{\mu }=R_{\ \nu \alpha \beta }^{\mu }+%
\frac{1}{4}(g^{\mu \lambda }g^{kl}g_{\beta \lambda ,k}g_{\alpha \nu
,l}-g^{\mu \lambda }g^{kl}g_{\alpha \lambda ,k}g_{\beta \nu ,l}).  \label{7}
\end{equation}%
Similarly, we get the component of the Riemann tensor with Latin indices:%
\begin{equation}
\mathcal{R}_{\ jkl}^{i}=R_{\ jkl}^{i}+\frac{1}{4}(g^{\lambda \tau
}g^{im}g_{lm,\lambda }g_{kj,\tau }-g^{\lambda \tau }g^{im}g_{km,\lambda
}g_{lj,\tau }).  \label{8}
\end{equation}%
We also have%
\begin{equation}
\mathcal{R}_{\ i\nu j}^{\mu }=\partial _{\nu }\Gamma _{ji}^{\mu }-\partial
_{j}\Gamma _{\nu i}^{\mu }+\Gamma _{\nu \lambda }^{\mu }\Gamma
_{ji}^{\lambda }-\Gamma _{j\lambda }^{\mu }\Gamma _{\nu i}^{\lambda }+\Gamma
_{\nu k}^{\mu }\Gamma _{ji}^{k}-\Gamma _{jk}^{\mu }\Gamma _{\nu i}^{k}.
\label{9}
\end{equation}%
Defining the covariant derivatives $\mathcal{D}_{\nu }\Gamma _{ij}^{\mu
}=\partial _{\nu }\Gamma _{ij}^{\mu }+\Gamma _{\nu \lambda }^{\mu }\Gamma
_{ji}^{\lambda }$ and $\mathcal{D}_{j}\Gamma _{\nu i}^{\mu }=\partial
_{j}\Gamma _{\nu i}^{\mu }-\Gamma _{\nu k}^{\mu }\Gamma _{ji}^{k}$, the
expression (9) is reduced to%
\begin{equation}
\mathcal{R}_{\ i\nu j}^{\mu }=\mathcal{D}_{\nu }\Gamma _{ji}^{\mu }-\mathcal{%
D}_{j}\Gamma _{\nu i}^{\mu }-\Gamma _{j\lambda }^{\mu }\Gamma _{\nu
i}^{\lambda }-\Gamma _{jk}^{\mu }\Gamma _{\nu i}^{k}.  \label{10}
\end{equation}%
Thus, by using (4), we find

\begin{equation}
\mathcal{R}_{\ i\nu j}^{\mu }=-\frac{1}{2}\mathcal{D}_{\nu }g_{ij}^{\quad
,\mu }-\frac{1}{2}\mathcal{D}_{j}(g^{\mu \lambda }g_{\lambda \nu ,i})-\frac{1%
}{4}g^{\mu \lambda }g^{\alpha \sigma }g_{\lambda \alpha ,j}g_{\sigma \nu ,i}+%
\frac{1}{4}g^{kl}g_{kj}^{\quad ,\mu }g_{li,\nu }.  \label{11}
\end{equation}%
With a similar calculation, we obtain

\begin{equation}
\mathcal{R}_{\ \mu j\nu }^{i}=-\frac{1}{2}\mathcal{D}_{j}g_{\mu \nu }^{\quad
,i}-\frac{1}{2}\mathcal{D}_{\nu }(g^{il}g_{lj,\mu })-\frac{1}{4}%
g^{il}g^{km}g_{lk,\nu }g_{mj,\mu }+\frac{1}{4}g^{\alpha \lambda }g_{\alpha
\nu }^{\quad ,i}g_{\lambda \mu ,j}.  \label{12}
\end{equation}

From the Ricci tensor 
\begin{equation}
\mathcal{R}_{AB}=\mathcal{R}_{\ AKB}^{K},  \label{13}
\end{equation}%
we can construct the curvature scalar $\mathcal{R}$:%
\begin{equation}
\mathcal{R}=g^{\mu \nu }\mathcal{R}_{\mu \nu }+g^{ij}\mathcal{R}_{ij}.
\label{14}
\end{equation}%
We can explicitly rewrite (14) as follows:%
\begin{equation}
\mathcal{R}=g^{\mu \nu }(\mathcal{R}_{\ \mu \alpha \nu }^{\alpha }+\mathcal{R%
}_{\ \mu k\nu }^{k})+g^{ij}(\mathcal{R}_{\ i\alpha j}^{\alpha }+\mathcal{R}%
_{\ ikj}^{k}).  \label{15}
\end{equation}%
Using the symmetric property $g^{ij}\mathcal{R}_{\ i\alpha j}^{\alpha
}=g^{\mu \nu }\mathcal{R}_{\ \mu k\nu }^{k}$, (15) can be simplified to 
\begin{equation}
\mathcal{R}=g^{\mu \nu }\mathcal{R}_{\ \mu \alpha \nu }^{\alpha }+g^{ij}%
\mathcal{R}_{\ ikj}^{k}+2g^{\mu \nu }g^{ij}\mathcal{R}_{\mu i\nu j}.
\label{16}
\end{equation}

Considering (7), (8), (11) and (12), and writing as $^{1}R=g^{\mu \nu }R_{\
\mu \alpha \nu }^{\alpha }$ and $^{2}R=g^{ij}R_{\ ikj}^{k}$ the Ricci scalar
in $p$ and $q$ dimensions, respectively, (16) leads to

\begin{equation}
\begin{array}{c}
\mathcal{R}=\ ^{1}R+\ ^{2}R \\ 
\\ 
-\frac{1}{4}g^{\mu \nu }g^{\alpha \beta }g^{ij}g_{\mu \nu ,i}g_{\alpha \beta
,j}+\frac{1}{4}g^{\mu \nu }g^{\alpha \beta }g^{ij}g_{\mu \alpha ,i}g_{\nu
\beta ,j} \\ 
\\ 
-\frac{1}{4}g^{\mu \nu }g^{ij}g^{kl}g_{ij,\mu }g_{kl,\nu }+\frac{1}{4}g^{\mu
\nu }g^{ij}g^{kl}g_{ik,\mu }g_{jl,\nu } \\ 
\\ 
-\frac{1}{2}g^{ij}\mathcal{D}_{\mu }g_{ij}^{\quad ,\mu }-\frac{1}{2}\mathcal{%
D}_{i}(g^{\mu \nu }g_{\mu \nu }^{\quad ,i}) \\ 
\\ 
-\frac{1}{4}g^{\mu \nu }g^{\alpha \beta }g^{ij}g_{\mu \alpha ,i}g_{\nu \beta
,j}+\frac{1}{4}g^{\mu \nu }g^{ij}g^{kl}g_{ik,\mu }g_{jl,\nu } \\ 
\\ 
-\frac{1}{2}g^{\mu \nu }\mathcal{D}_{i}g_{\mu \nu }^{\quad ,i}-\frac{1}{2}%
\mathcal{D}_{\mu }(g^{ij}g_{ij}^{\quad ,\mu }) \\ 
\\ 
-\frac{1}{4}g^{\mu \nu }g^{ij}g^{kl}g_{ik,\mu }g_{jl,\nu }+\frac{1}{4}g^{\mu
\nu }g^{\alpha \beta }g^{ij}g_{\mu \alpha ,i}^{\quad }g_{\nu \beta ,j}.%
\end{array}
\label{17}
\end{equation}%
This expression can be simplified to%
\begin{equation}
\begin{array}{c}
\mathcal{R}=\ ^{1}R+\ ^{2}R-\mathcal{D}_{\mu }(g^{ij}g_{ij}^{\quad ,\mu })-%
\mathcal{D}_{i}(g^{\mu \nu }g_{\mu \nu }^{\quad ,i}) \\ 
\\ 
-\frac{1}{4}g^{\mu \nu }g^{\alpha \beta }g^{ij}g_{\mu \nu ,i}g_{\alpha \beta
,j}-\frac{1}{4}g^{\mu \nu }g^{\alpha \beta }g^{ij}g_{\mu \alpha ,i}g_{\nu
\beta ,j} \\ 
\\ 
-\frac{1}{4}g^{\mu \nu }g^{ij}g^{kl}g_{ij,\mu }g_{kl,\nu }-\frac{1}{4}g^{\mu
\nu }g^{ij}g^{kl}g_{ik,\mu }g_{jl,\nu }.%
\end{array}
\label{18}
\end{equation}

\bigskip Therefore, we obtain the action 
\begin{equation}
\begin{array}{c}
\mathcal{S}=\int_{M}\sqrt{^{1}g}\sqrt{^{2}g}\mathcal{R}=\int_{M}[\sqrt{^{1}g}%
\sqrt{^{2}g}\ ^{1}R+\sqrt{^{1}g}\sqrt{^{2}g}\ ^{2}R \\ 
\\ 
-\sqrt{^{1}g}\mathcal{D}_{\mu }(\sqrt{^{2}g}g^{ij}g_{ij}^{\quad ,\mu })-%
\sqrt{^{2}g}\mathcal{D}_{i}(\sqrt{^{1}g}g^{\mu \nu }g_{\mu \nu }^{\quad ,i})
\\ 
\\ 
+\frac{1}{4}\sqrt{^{1}g}\sqrt{^{2}g}(g^{\mu \nu }g^{\alpha \beta
}g^{ij}g_{\mu \nu ,i}g_{\alpha \beta ,j}-g^{\mu \nu }g^{\alpha \beta
}g^{ij}g_{\mu \alpha ,i}g_{\nu \beta ,j} \\ 
\\ 
+g^{\mu \nu }g^{ij}g^{kl}g_{ij,\mu }g_{kl,\nu }-g^{\mu \nu
}g^{ij}g^{kl}g_{ik,\mu }g_{jl,\nu })].%
\end{array}
\label{19}
\end{equation}%
It is reasonable to assume that at large distances $\int_{M}\sqrt{^{1}g}%
\mathcal{D}_{\mu }(\sqrt{^{2}g}g^{ij}g_{ij}^{\quad ,\mu })=0$ and $\int_{M}%
\sqrt{^{2}g}\mathcal{D}_{i}(\sqrt{^{1}g}g^{\mu \nu }g_{\mu \nu }^{\quad
,i})=0$. Therefore (19) can be simplified as follows [17]%
\begin{equation}
\begin{array}{c}
\mathcal{S}=\int_{M}\sqrt{^{1}g}\sqrt{^{2}g}\mathcal{R}=\int_{M}\sqrt{^{1}g}%
\sqrt{^{2}g}\ [^{1}R+^{2}R+ \\ 
\\ 
+\frac{1}{4}g^{\mu \nu }g^{\alpha \beta }g^{ij}(g_{\mu \nu ,i}g_{\alpha
\beta ,j}-g_{\mu \alpha ,i}g_{\nu \beta ,j})+\frac{1}{4}g^{\mu \nu
}g^{ij}g^{kl}(g_{ij,\mu }g_{kl,\nu }-g_{ik,\mu }g_{jl,\nu })].%
\end{array}
\label{20}
\end{equation}%
Thus, we have achieved our first goal of splitting the action in terms of
the $M^{p}$ and $M^{q}$ spaces. However, observe that the last two terms in
(20) are interacting terms between the two metric fields $g_{ij}$ and $%
g_{\alpha \beta }$.\bigskip 

As an application of (20), we shall now discuss several examples. First, one
may assume that $g_{\mu \nu }=g_{\mu \nu }(x)$ and $g_{ij}=g_{ij}(y)$. In
this case $\mathcal{S}$ is reduced to%
\begin{equation}
\mathcal{S=}\int_{M}\sqrt{^{1}g}\sqrt{^{2}g}\ (^{1}R+^{2}R),  \label{21}
\end{equation}%
which is a well known result.\bigskip

More interesting cases may arise if we assume that $g_{\mu \nu }=\psi
^{2}(x,y)\tilde{g}_{\mu \nu }(x)$ and $g_{ij}=\varphi ^{2}(x,y)\tilde{g}%
_{ij}(y)$. Let us first substitute this assumption in the last two\ terms of
(20): 
\begin{equation}
\begin{array}{c}
\mathcal{S}=\int_{M}\sqrt{^{1}g}\sqrt{^{2}g}\ [^{1}R+^{2}R+ \\ 
\\ 
+p(p-1)\psi ^{-2}\varphi ^{-2}\tilde{g}^{ij}\psi _{,i}\psi _{,j}+q(q-1)\psi
^{-2}\varphi ^{-2}\tilde{g}^{\mu \nu }\varphi _{,\mu }\varphi _{,\nu }].%
\end{array}
\label{22}
\end{equation}%
Since the Ricci scalars $^{1}R$ and $^{2}R$ become

\begin{equation}
^{1}R=\psi ^{-2}[^{1}\tilde{R}-(p-1)(p-4)\psi ^{-2}\psi _{,\lambda }\psi
^{,\lambda }-2(p-1)\psi ^{-1}\nabla _{\lambda }\psi ^{,\lambda }]  \label{23}
\end{equation}%
and

\begin{equation}
^{2}R=\varphi ^{-2}[^{2}\tilde{R}-(q-1)(q-4)\varphi ^{-2}\varphi
_{,i}\varphi ^{,i}-2(q-1)\varphi ^{-1}\nabla _{i}\varphi ^{,i}]  \label{24}
\end{equation}%
respectively, after some rearrangements, we obtain\bigskip 
\begin{equation}
\begin{array}{c}
\mathcal{S=}\int_{M}\sqrt{^{1}\tilde{g}}\sqrt{^{2}\tilde{g}}[\psi
^{p-2}\varphi ^{q~1}\tilde{R}+\psi ^{p}\varphi ^{q-2~2}\tilde{R} \\ 
\\ 
-(p-1)(p-4)\psi ^{p-4}\varphi ^{q}\psi _{,\lambda }\psi ^{,\lambda
}-2(p-1)\psi ^{p-3}\varphi ^{q}\nabla _{\lambda }\psi ^{,\lambda } \\ 
\\ 
+p(p-1)\psi ^{p-2}\varphi ^{q-2}\psi _{,i}\psi ^{,i}\} \\ 
\\ 
-(q-1)(q-4)\psi ^{p}\varphi ^{q-4}\varphi _{,i}\varphi ^{,i}-2(q-1)\psi
^{p}\varphi ^{q-3}\nabla _{i}\varphi ^{,i} \\ 
\\ 
+q(q-1)\psi ^{p-2}\varphi ^{q-2}\varphi _{,\lambda }\varphi ^{,\lambda }\}].%
\end{array}
\label{25}
\end{equation}

\bigskip Integrating by parts, (25) yields

\begin{equation}
\begin{array}{c}
\mathcal{S=}\int_{M}\sqrt{^{1}\tilde{g}}\sqrt{^{2}\tilde{g}}[(\psi
^{p-2}\varphi ^{q})^{~1}\tilde{R}+(\psi ^{p}\varphi ^{q-2})^{~2}\tilde{R} \\ 
\\ 
+(p-1)(p-2)\psi ^{p-4}\varphi ^{q}\psi _{,\lambda }\psi ^{,\lambda
}+2q(p-1)\psi ^{p-3}\varphi ^{q-1}\varphi _{,\lambda }\psi ^{,\lambda } \\ 
\\ 
+p(p-1)\psi ^{p-2}\varphi ^{q-2}\psi _{,i}\psi ^{,i} \\ 
\\ 
+(q-1)(q-2)\psi ^{p}\varphi ^{q-4}\varphi _{,i}\varphi ^{,i}+2p(q-1)\psi
^{p-1}\varphi ^{q-3}\varphi _{,i}\psi ^{,i} \\ 
\\ 
+q(q-1)\psi ^{p-2}\varphi ^{q-2}\varphi _{,\lambda }\varphi ^{,\lambda }].%
\end{array}
\label{26}
\end{equation}

Here, we are interested in exploring a possible duality symmetry in (26).
For this purpose, let us consider the special case $\psi =\varphi ^{-1}$. We
have%
\begin{equation}
\begin{array}{c}
\mathcal{S=}\int_{M}\sqrt{^{1}\tilde{g}}\sqrt{^{2}\tilde{g}}[(\varphi
^{q-p+2})^{~1}\tilde{R}+(\varphi ^{q-p-2})^{~2}\tilde{R} \\ 
\\ 
+(p-1)(p-2)\varphi ^{q-p}\varphi _{,\lambda }\varphi ^{,\lambda
}+2q(p-1)\varphi ^{q-p}\varphi _{,\lambda }\varphi ^{,\lambda } \\ 
\\ 
+p(p-1)\varphi ^{q-p-4}\varphi _{,i}\varphi ^{,i} \\ 
\\ 
+(q-1)(q-2)\varphi ^{q-p-4}\varphi _{,i}\varphi ^{,i}+2p(q-1)\varphi
^{q-p-4}\varphi _{,i}\varphi ^{,i} \\ 
\\ 
+q(q-1)\varphi ^{q-p}\varphi _{,\lambda }\varphi ^{,\lambda }],%
\end{array}
\label{27}
\end{equation}%
which can also be rewritten as

\begin{equation}
\begin{array}{c}
\mathcal{S=}\int_{M}\sqrt{^{1}\tilde{g}}\sqrt{^{2}\tilde{g}}[(\varphi
^{q-p+2})^{~1}\tilde{R}+(\varphi ^{q-p-2})^{~2}\tilde{R} \\ 
\\ 
+[(p-1)(p-2)+2q(p-1)+q(q-1)]\varphi ^{q-p}\varphi _{,\lambda }\varphi
^{,\lambda } \\ 
\\ 
+[(q-1)(q-2)+2p(q-1)+p(p-1)]\varphi ^{q-p-4}\varphi _{,i}\varphi ^{,i}].%
\end{array}
\label{28}
\end{equation}

\bigskip Since%
\begin{equation}
(p-1)(p-2)+2q(p-1)+q(q-1)=(p+q-1)(p+q-2)  \label{29}
\end{equation}%
and%
\begin{equation}
(q-1)(q-2)+2p(q-1)+p(p-1)=(p+q-1)(p+q-2),  \label{30}
\end{equation}%
we can further simplify (28) to\bigskip 
\begin{equation}
\mathcal{S=S}_{p}+\mathcal{S}_{q},  \label{31}
\end{equation}%
where%
\begin{equation}
\mathcal{S}_{p}=\int_{M}\sqrt{^{1}\tilde{g}}\sqrt{^{2}\tilde{g}}[(\varphi
^{q-p+2})^{~1}\tilde{R}+(p+q-1)(p+q-2)\varphi ^{q-p}\varphi _{,\lambda
}\varphi ^{,\lambda }]  \label{32}
\end{equation}%
and%
\begin{equation}
\mathcal{S}_{q}=\int_{M}\sqrt{^{1}\tilde{g}}\sqrt{^{2}\tilde{g}}[(\varphi
^{q-p-2})^{~2}\tilde{R}+(p+q-1)(p+q-2)\varphi ^{q-p-4}\varphi _{,i}\varphi
^{,i}].  \label{33}
\end{equation}%
We are interested in the possible invariance of the action (31) under the
duality transformation $\varphi \rightarrow \varphi ^{-1}$. Applying this
transformation to (31), we obtain

\begin{equation}
\mathcal{S}_{p}\rightarrow \int_{M}\sqrt{^{1}\tilde{g}}\sqrt{^{2}\tilde{g}}%
[\varphi ^{p-q-2}{}^{~1}\tilde{R}+(p+q-1)(p+q-2)\varphi ^{p-q-4}\varphi
_{,\lambda }\varphi ^{,\lambda }],  \label{34}
\end{equation}%
while%
\begin{equation}
\mathcal{S}_{q}\rightarrow \int_{M}\sqrt{^{1}\tilde{g}}\sqrt{^{2}\tilde{g}}%
[\varphi ^{p-q+2}{}^{~2}\tilde{R}+(p+q-1)(p+q-2)\varphi ^{p-q}\varphi
_{,i}\varphi ^{,i}].  \label{35}
\end{equation}%
Therefore we observe that if $p\rightarrow q+2$ and $q\rightarrow p-2$, then 
$\mathcal{S}_{p}$ remains invariant, but%
\begin{equation}
\mathcal{S}_{q}=\int_{M}\sqrt{^{1}\tilde{g}}\sqrt{^{2}\tilde{g}}[\varphi
^{p-q+6}{}^{~2}\tilde{R}+(p+q-1)(p+q-2)\varphi ^{p-q+4}\varphi _{,i}\varphi
^{,i}];  \label{36}
\end{equation}%
that is, $\mathcal{S}_{q}$ is not invariant. Conversely, if $p\rightarrow
q-2 $ and $q\rightarrow p+2$, then $\mathcal{S}_{q}$ is invariant, while $%
\mathcal{S}_{p}$ becomes 
\begin{equation}
\mathcal{S}_{p}=\int_{M}\sqrt{^{1}\tilde{g}}\sqrt{^{2}\tilde{g}}[\varphi
^{p-q-6}{}^{~1}\tilde{R}+(p+q-1)(p+q-2)\varphi ^{p-q-4}\varphi _{,\lambda
}\varphi ^{,\lambda }],  \label{37}
\end{equation}%
which means that $\mathcal{S}_{p}$ is not invariant. Therefore, we have the
peculiar situation that under the duality transformation $\varphi
\rightarrow \varphi ^{-1}$, $\mathcal{S}_{p}$ or $\mathcal{S}_{q}$ remains
invariant, but not both, depending on having $p\rightarrow q+2$ and $%
q\rightarrow p-2$, (that is $p\rightleftarrows q+2$), or $p\rightarrow q-2$
and $q\rightarrow p+2$ ( $p\rightleftarrows q-2$) , respectively. \bigskip

This duality property for the action (31) resembles the Farkas property in
oriented matroid theory. In order to fully appreciate this comment, let us
explain briefly what the Farkas property means. First, let us consider a
total tangent bundle $T=(H,V)$ associated with $M$, where $H$ and $V$ denote
the horizontal and vertical parts of $T$. Assume that $L\subseteq M$
corresponds to $H$ and \ that the orthogonal complement $L_{\perp }$
corresponds to $V$. Now, just as $(H,V)$ determine the structure of $T$, the
dual pair $(L,L_{\perp })$ determines the structure of the total space $M$.
It turns out that one can introduce the concept of an oriented matroid in
terms of the structure $(L,L_{\perp })$, rather than only in terms of the
subspace $L$. (For details in oriented matroids, see Ref. [6].) One can
prove that a transition of the form%
\begin{equation}
(L,L_{\perp })\rightarrow (H,V),  \label{38}
\end{equation}%
ensures a duality symmetry, which is one of the main subjects of oriented
matroid theory. In fact, there exist a formal definition of an oriented
matroid in terms of the analogue of $(L,L_{\perp })$. Such a definition uses
the concept of Farkas property, which we shall now proceed to discuss
briefly (see Ref. [5] for details).\bigskip

Let us first describe the sign vector concept. Let $E\neq \varnothing $ be a
finite set. An element $X\in \{+,-,0\}^{E}$ is called a sign vector. The
positive, negative and zero parts of $X$ are denoted by $X^{+}$, $X^{-}$ and 
$X^{0}$ respectively. Further, we define $suppX\equiv X^{+}\cup X^{-}$.
Consider two sets $S$ and $S^{\prime }$ of signed vectors. The pair $%
(S,S^{\prime })$ is said to have the Farkas property, if $\forall e\in E$
either\bigskip

(Fa)$\exists X\in S$, $e\in suppX$ and $X\geq 0$

or

(Fb) $\exists Y\in S^{\prime }$, $e\in suppY$ and $Y\geq 0$,\bigskip

\noindent but not both. Here, $X\geq 0$ means that $X$ has a positive (+) or
a zero (0) entry in every coordinate. Observe that $(S,S^{\prime })$ has the
Farkas property if and only if $(S^{\prime },S)$ has it. Let $S$ be a set of
signed vectors, and let $I$ and $J$ denote disjoint subsets of $E$. Then%
\begin{equation}
S\backslash I/J=\{\tilde{X}|\exists X\in S,X_{I}=0,X_{J}=\ast ,X=\tilde{X}%
\text{ on }E\backslash (I\cup J)\},  \label{39}
\end{equation}%
is called a minor of $S$ (obtained by deleting $I$ and contracting $J$).
Here, the symbol "$\ast $"\ denotes and arbitrary value. If $S$ and $%
S^{\prime }$ are sets of sign vectors on $E$, then $(S\backslash I/J,S\prime
\backslash J/I)$ is called minor of $(S,S^{\prime })$. Similarly,%
\begin{equation}
_{I}S=\{\tilde{X}|\exists X\in S,X_{I}=-\tilde{X}_{I},X_{E\backslash I}=%
\tilde{X}_{E\backslash I}\}  \label{40}
\end{equation}%
is called the reorientation of $S$ on $I$. Further, $(_{I}S,_{I}S^{\prime })$
is the reorientation of $(S,S^{\prime })$ on $I$. Moreover, $S$ is symmetric
if $S=-S$, where $-S$ is the set of signed vectors which are opposite to the
signed vectors of $S$.\bigskip

We can now give a definition of oriented matroids in terms of the Farkas
property. Let $E\neq \varnothing $ be a finite set and let $S$ and $%
S^{\prime }$ two sets of sign vectors. The pair $(S,S^{\prime })$ is called
an oriented matroid on $E$, if\bigskip

(O1) $S$ and $S^{\prime }$ are symmetric, and

(O2) every reorientation of every minor of $(S,S^{\prime })$ has the Farkas
property.\bigskip

\noindent From this definition it follows that $(S^{\prime },S)$ is also an
oriented matroid as it is every reorientation and every minor of $%
(S,S^{\prime })$.\bigskip

Two sign vectors $X$ and $Y$ are orthogonal if%
\begin{equation}
(X^{+}\cap Y^{+})\cup (X^{-}\cap Y^{-})\neq \varnothing \Leftrightarrow
(X^{+}\cap Y^{-})\cup (X^{-}\cap Y^{+})\neq \varnothing .  \label{41}
\end{equation}%
Accordingly, we denote the orthogonal complement of $S$ by $S_{\perp }$, and
it is defined by%
\begin{equation}
S_{\perp }=\{Y|Y\perp X\text{ for all }X\in S\}.  \label{42}
\end{equation}%
If $S^{\prime }\subseteq S_{\perp }$, then $S$ and $S^{\prime }$ can be
considered as orthogonal. \bigskip

Coming back to our problem at hand we have that, due to the form of the
metric (1), $\mathcal{S}_{p}$ and $\mathcal{S}_{q}$ can be associated with
the two orthogonal subspaces $M_{p}$ and $M_{q}$ of $M$, respectively. This
suggests to introduce the analogue of the Farkas property for the action
(31):\bigskip\ For every transformation $\varphi \rightarrow \varphi ^{-1}$,

(Ha) $\exists p$ for $M_{p}$, such that $\mathcal{S}_{p}$ is invariant

or

(Hb) $\exists q$ for $M_{q}$, such that $\mathcal{S}_{q}$ is
invariant,\bigskip

\noindent but not both.\bigskip

It is interesting to write (31) in the alternative form

\begin{equation}
\begin{array}{c}
\mathcal{S=}\int_{M}\sqrt{^{1}\tilde{g}}\sqrt{^{2}\tilde{g}}[\varphi
^{q-p}(\gamma ^{\mu \nu }\tilde{R}_{\mu \nu }+\gamma ^{ij}\tilde{R}_{ij}) \\ 
\\ 
+(d-1)(d-2)\varphi ^{q-p-2}(\gamma ^{\lambda \sigma }\varphi _{,\lambda
}\varphi _{,\sigma }+\gamma ^{ij}\varphi _{,i}\varphi _{,j})].%
\end{array}
\label{43}
\end{equation}%
Here, we considered the fact that $^{1}\tilde{R}=\tilde{g}^{\mu \nu }\tilde{R%
}_{\mu \nu }$, $^{2}\tilde{R}=\tilde{g}^{ij}\tilde{R}_{ij}$, $\gamma ^{\mu
\nu }=\varphi ^{2}(x,y)\tilde{g}^{\mu \nu }(x)$ and $\gamma ^{ij}=\varphi
^{-2}(x,y)\tilde{g}^{ij}(x)$. This expression can be written in the more
compact form

\begin{equation}
\mathcal{S=}\int_{M}\sqrt{^{1}\tilde{g}}\sqrt{^{2}\tilde{g}}\varphi
^{q-p}[\gamma ^{AB}\tilde{R}_{AB}+(d-1)(d-2)\varphi ^{-2}\gamma ^{AB}\varphi
_{,A}\varphi _{,B}].  \label{44}
\end{equation}

Let us make some final comments. Usually, one is interested in the
invariance of an action $\mathcal{S}$ under certain infinitesimal
transformations. Here we have shown that, if an action can be divided in two
complementary "orthogonal" actions $\mathcal{S}_{p}$ and $\mathcal{S}_{q}$,
then the invariance of the total action $\mathcal{S=S}_{p}+\mathcal{S}_{q}$
under duality transformations is not what really matters, but rather whether 
$\mathcal{S}_{p}$ or $\mathcal{S}_{q}$ are invariant, but not both, as the
analogue of the Farkas property should require.\ Since the results obtained
above are valid for arbitrary dimensionality -albeit the conditions imposed
for the metric-, they can be of interest for cosmology with extra
dimensions, in particular when those extra dimensions are non-compact
[14][15]. Also, since \textit{a priori} it has been not chosen any signature
for the metric $\gamma _{AB}$, our analysis may give some insight in 2t
physics [16].

\bigskip

\ 

\begin{center}
\textbf{Acknowledgments}

\bigskip
\end{center}

This work was partially supported by PROFAPI 2008 and 2009.

\bigskip\

\end{document}